\documentclass[aps,prl,reprint,groupedaddress,floatfix,superscriptaddress,amsmath]{revtex4-1}

\makeatletter

\usepackage{changes}
\usepackage{graphicx}

\makeatother

\begin{document}
\title{$\pi$ phase interlayer shift and stacking fault in the kagome superconductor CsV$_{3}$Sb$_{5}$}

\author{Feng\,Jin}
\email[e-mail:]{jinfeng@iphy.ac.cn}
\affiliation{Beijing National Laboratory for Condensed Matter Physics, Institute of Physics, Chinese Academy of Sciences, Beijing 100190, China}

\author{Wei\,Ren}
\affiliation{School of Physical Science and Technology, Lanzhou University, Lanzhou 730000, China}

\author{Mingshu\,Tan}
\affiliation{School of Physical Science and Technology, Lanzhou University, Lanzhou 730000, China}

\author{Mingtai\,Xie}
\affiliation{School of Physical Science and Technology, Lanzhou University, Lanzhou 730000, China}

\author{Bingru\,Lu}
\affiliation{School of Physical Science and Technology, Lanzhou University, Lanzhou 730000, China}

\author{Zheng\,Zhang}
\affiliation{Beijing National Laboratory for Condensed Matter Physics, Institute of Physics, Chinese Academy of Sciences, Beijing 100190, China}

\author{Jianting\,Ji}
\affiliation{Beijing National Laboratory for Condensed Matter Physics, Institute of Physics, Chinese Academy of Sciences, Beijing 100190, China}

\author{Qingming\,Zhang}
\email[e-mail:]{qmzhang@iphy.ac.cn}
\affiliation{School of Physical Science and Technology, Lanzhou University, Lanzhou 730000, China}
\affiliation{Beijing National Laboratory for Condensed Matter Physics, Institute of Physics, Chinese Academy of Sciences, Beijing 100190, China}

\begin{abstract}
The stacking degree of freedom is a crucial factor in tuning material properties and has been extensively investigated in layered materials. The kagome superconductor CsV$_{3}$Sb$_{5}$ was recently discovered to exhibit a three-dimensional CDW phase below $T_{CDW} \sim$ 94 K. Despite the thorough investigation of in-plane modulation, the out-of-plane modulation has remained ambiguous. Here, our polarization- and temperature-dependent Raman measurements reveal the breaking of $C_6$ rotational symmetry and the presence of three distinct domains oriented at approximately 120 degrees to each other. The observations demonstrate that the CDW phase can be naturally explained as a 2c staggered order phase with adjacent layers exhibiting a relative $\pi$ phase shift. Further, we discover a first-order structural phase transition at approximately 65 K and suggest that it is a stacking order-disorder phase transition due to stacking fault, supported by the thermal hysteresis behavior of a Cs-related phonon mode. Our findings highlight the significance of the stacking degree of freedom in CsV$_{3}$Sb$_{5}$ and offer structural insights to comprehend the entanglement between superconductivity and CDW.
\end{abstract}

\maketitle

Exploring the emergent physics from simple lattice models plays a vital role in modern condensed matter physics~\cite{Onsager1944}. The kagome lattice, made of corner-sharing triangles, has attracted tremendous attention since it was first introduced by Sy\^ozi~\cite{Syozi1951}. In the context of frustrated magnetism, the insulating kagome lattice with antiferromagnetic exchange interactions exhibits no phase transition, making it a promising candidate for the long-sought quantum spin liquid states~\cite{Wulferding2010, Han2012, Fu2015}. In the absence of magnetism, the metallic kagome lattice features nontrivial topological band structures (Dirac fermions, flat bands and van Hove singularities), and can support various correlated phenomena, including charge density wave (CDW), superconductivity, charge bond order and electronic nematicity~\cite{Ortiz2019,Li2021,Ortiz2020,Guo2009,Nie2022}. Because of its elegant structure and these exotic quantum phenomena, the kagome lattice has been intensely studied in condensed matter physics.

Recently, layered kagome metals AV$_{3}$Sb$_{5}$(A = K, Rb, Cs) have aroused tremendous research interest, owing to possible unconventional superconductivity, exotic CDW ordering and nematicity~\cite{Ortiz2021,Yin2021,Oey2022,Zhang2021,Zhu2022,Li2022,Chen2021,Chen2022,Ratcliff2021}. Among the family members, CsV$_{3}$Sb$_{5}$ undergoes a three-dimensional CDW transition at $T_{CDW} \sim$ 94 K with both in-plane and out-of-plane modulations, which breaks the $C_6$ rotational symmetry and the time reversal symmetry~\cite{ Xiang2021, Zhao2021, Xu2022, Mielke2022, Jiang2021}. The in-plane 2 $\times$ 2 modulation has either a star-of-David (SD) or the inverse star of David (ISD) pattern~\cite{Tan2021,Christensen2021}, both of which preserve the $C_6$ rotational symmetry. Thus, the out-of-plane modulation or the stacking configuration determines how the $C_6$ rotational symmetry is broken. However, the stacking issue is little explored despite that its importance had been recognized by many experiments~\cite{Zheng2022,Chen2021a,Yu2021,Ortiz20212,Stahl2022}.

More importantly, recent progress has shown that the stacking configuration plays an important role in both the CDW transition and the superconducting transition. For example, it was reported that the CDW transition at 94 K is of order-disorder type~\cite{Subires2023} and there is a competition between 2c and 4c CDW phases below $T_{CDW}$~\cite{Xiao2023}. These experiments hint that the state below $T_{CDW}$ is a stacking disorder state. Furthermore, it seems that not only the CDW transition but also the superconducting transition strongly depends on the interlayer interaction. For instance,  the superconducting transition temperature exhibits a two-dome-like behavior with applying pressure~\cite{Zheng2022,Chen2021a,Yu2021} and a nonmonotonic behavior with reducing sample thickness~\cite{Song2023}. These experiments suggest that the stacking order serves as a tuning knob to alter both the CDW transition and the superconducting transition. Hence, the knowledge about it is highly needed to understand these unconventional phenomena and the entanglement between them.

\begin{figure*}[t]
\centering
\includegraphics[width=17cm]{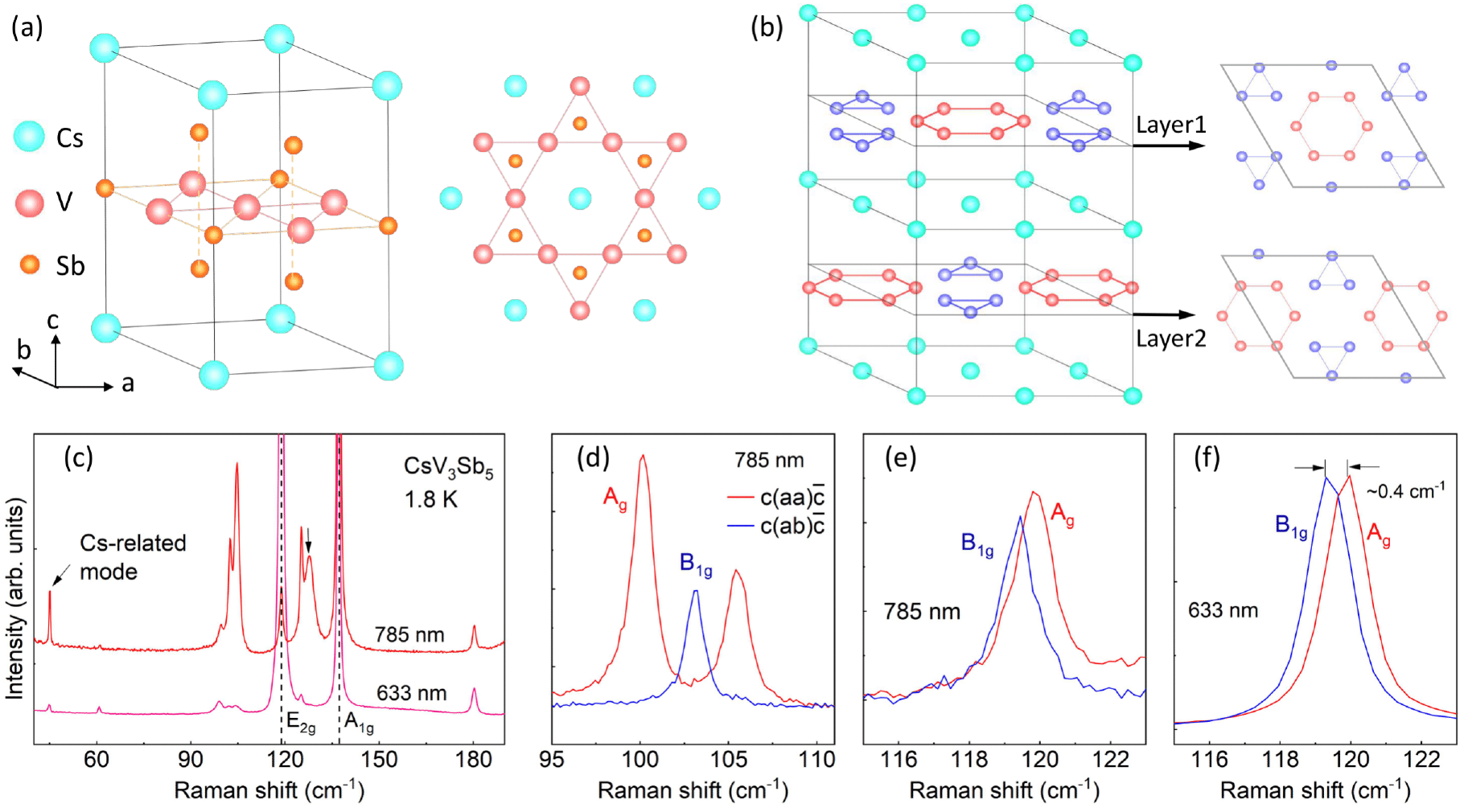}
\caption{(a) Crystal structure of CsV$_{3}$Sb$_{5}$ in the pristine phase (left) and its top view (right). (b) Illustration of the crystal structure in the 2c staggered CDW phase in which there is a relative $\pi$ phase shift between neighboring layers (right). (c) Raman spectra of CsV$_{3}$Sb$_{5}$ taken at 1.8 K with 785- and 633-nm lasers. The dashed lines show the positions of the E$_{2g}$ and A$_{1g}$ modes in the pristine phase. (d-f) Polarized Raman spectra, collected with collinear and crossed configurations, in which two phonon modes can be resolved at around 119 cm$^{-1}$. Note that the splitting can only be resolved after determining the in-plane a- or b-axis (see Fig.~2) and collect Raman spectra with exactly c(aa)$\overline{\text{c}}$ and c(ab)$\overline{\text{c}}$ configurations.}
\end{figure*}

In this Letter, we performed polarization- and temperature-dependent Raman scattering measurements on CsV$_{3}$Sb$_{5}$ to explore the stacking configuration in its CDW phase. The splitting of the E$_{2g}$ phonon and the anisotropic polarization-dependent behavior of phonon modes confirm the breaking of $C_6$ rotational symmetry. Additionally, three distinct domains that are oriented at $\sim$120$^{\circ}$ to each other can clearly be observed. We propose that a $\pi$ phase shift between neighboring layers can explain both the $C_6$ rotational symmetry breaking and the three distinct domains. Further, we discover a first-order structural phase transition at approximately 65 K and propose that it is a stacking order-disorder phase transition due to stacking fault, as evidenced by the thermal hysteresis behavior of a Cs-related phonon mode.

 \begin{figure*}[t]
\centering 
\includegraphics[width=17cm]{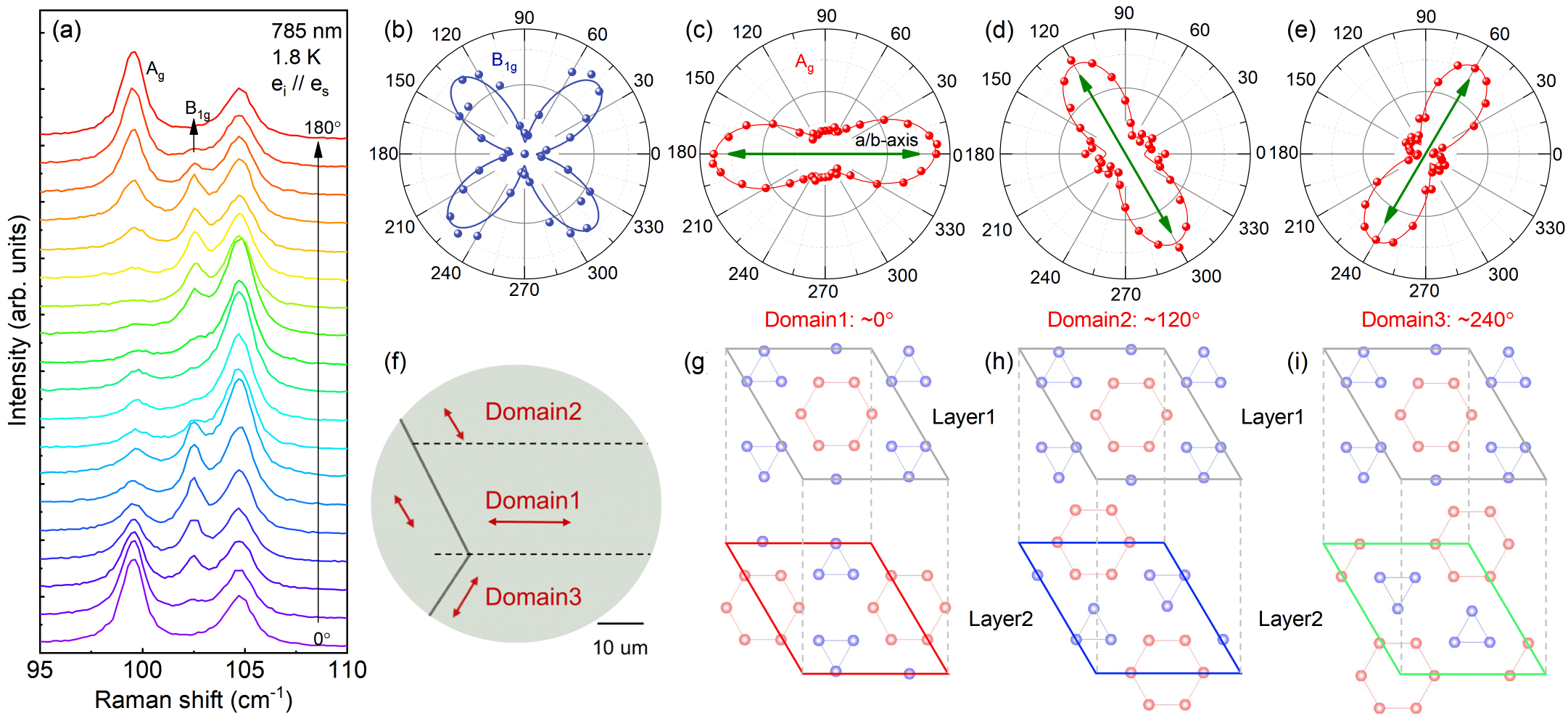}
\caption{(a) Polarization angle dependent Raman spectra, collected with parallel polarization configuration. (b-c) The anisotropic behavior of the A$g$ and B$_{1g}$ modes which can be well described by the equation (1). (d-e) Same as panel (c) but measured in other two domains (see raw spectra in Ref.~\cite{SuppleM}). The arrows indicate the direction of the in-plane a/b-axis, which are rotated by $\pm$120$^{\circ}$ with respected to each other. (f) Schematic illustration of three different domains (see optical image in Ref.~\cite{SuppleM}). The three domains are randomly distributed depending on the edges (solid lines) and/or impurities (dashed lines). (g-i) Schematic illustrations about the formation of the three domains. Compared with Layer1, the V-hexagon at Layer2 has a $\pi$ phase shift along six different directions, resulting in the formation of three different domains.}
\end{figure*}

The CsV$_{3}$Sb$_{5}$ single crystal used in this study was grown via the self-flux method~\cite{Ortiz2019} and has been characterized before Raman measurements. The as-grown sample was cleaved in the air to expose a fresh (001) crystallographic plane. Then, the cleaved crystal was transferred into an AttoDRY 2100 cryostat which utilizes helium as the exchange gas and allows for cooling down to 1.8 K.  Below 120 K, the cooling rate is less than 1.5 K/min and thus the cooling process can be considered as a quasistatic one~\cite{SuppleM}. Confocal micro-Raman measurements were performed with a backscattering configuration using a Jobin Yvon HR-Evolution system consisting of two laser lines: a 632.8-nm HeNe laser and a diode-pumped solid-state 785-nm laser. The excitation laser beam was focused into a spot of $\sim$5 $\mu$m in diameter on the ab-plane with a power of $<$100 $\mu$W to avoid overheating.

The layered kagome metal CsV$_{3}$Sb$_{5}$ crystallizes into the P6/mmm ($D_{6h}^1$) space group with alternating layers of $V_3Sb_5$ and Cs (Fig.~1a). The V atoms in the $V_3Sb_5$ layer form a perfect kagome lattice whereas the Cs atoms form a triangular lattice sandwiched by two $V_3Sb_5$ layers. As a result, the pristine lattice has $C_6$ rotational symmetry. Below $T_{CDW}$, the V-kagome layer shows an in-plane 2 $\times$ 2 modulation, forming either a SD or the ISD pattern, with $C_6$ rotational symmetry remained in each layer. However, if there is a finite interaction between layers and a $\pi$ phase shift exists between neighboring layers, the $C_6$ rotational symmetry will reduce to $C_2$ symmetry and the structure will have the $D_{2h}$ point group (Fig.~1b).

For Raman measurements, a direct way to confirm the breaking of $C_6$ rotational symmetry is to examine the splitting of the E$_{2g}$ mode in the CDW phase. For this, polarized Raman scattering measurements are performed with a 785 nm laser (Fig.~1c), with which the spectra gain a high resolution. Compared to the spectra excited with 633 nm laser, the ones obtained with 785 nm laser exhibit several features: (1) a new mode located at $\sim$128 cm$^{-1}$ can be distinguished; (2) the Cs-related mode at $\sim$ 45 cm$^{-1}$ has a higher intensity; and (3) the three CDW-induced modes at around 100 cm$^{-1}$ can be well resolved (Fig.~1d). Importantly, the splitting of the E$_{2g}$ mode can be well resolved (Fig.~1e and 1f), which comfirms the breaking of $C_6$ rotational symmetry.

Another way to confirm the breaking of $C_6$ rotational symmetry is through polarization angle dependent measurements. We employ the collinear polarization configuration, in which the polarizations of the incident and scattered light remain parallel ($\hat{\textbf{e}}_i\parallel \hat{\textbf{e}}_s$) while rotating the sample around the c-axis. For the $D_{6h}$ point group, the Raman intensities of the A$_{1g}$ and E$_{2g}$ modes should exhibit no angle dependence with $I_{A_{1g}} \propto a^2$ and $I_{E_{2g}} \propto f^2$. However, for the $D_{2h}$ point group, the Raman tensors of the A$_{g}$ and B$_{1g}$ modes have the following forms,
 \begin{center}
$A_{g}=\left(
     \begin{array}{ccc}
       a&~~0~~&0\\
       0&~~b~~&0\\
       0&~~0~~&c\\
     \end{array}
     \right),~~~$
$B_{1g}=\left(
     \begin{array}{ccc}
       0&~~d~~&0\\
       d&~~0~~&0\\
       0&~~0~~&0\\
     \end{array}
     \right).$
 \end{center}
In this case, the Raman intensity of the A$_{g}$ and B$_{1g}$ mode has twofold symmetry and fourfold symmetry, respectively, with,
\begin{eqnarray}
 I_{A_{g}} \propto (a\cos^2\theta + b\sin^2\theta)^2;~~~~~  I_{B_{1g}} \propto d^2\sin^2(2\theta)
\end{eqnarray}

The polarization angle dependent spectra are shown in Fig.~2a. Thanks to the high resolution of the spectra, the modes around 100 cm$^{-1}$ can be well resolved and the intensity of them exhibits significant angle dependence, which distinguishes from the no angle-dependent behavior expected from the $D_{6h}$ point group. Furthermore, the extracted intensities of these two modes can be well fitted using the expected formula of the $D_{2h}$ point group (Fig.~2b and 2c), confirming that the low-temperature phase belongs to the $D_{2h}$ point group. Note that the anisotropy of the phonon mode cannot be ascribed to the superconductivity or the nematic order since it remains for temperature above the electronic transition temperatures~\cite{SuppleM}. These findings, including the splitting of the E$_{2g}$ mode and the anisotropic polarization-dependent behavior of the mode's intensities, confirm the breaking of $C_6$ rotational symmetry.

 \begin{figure}[t]
\centering
\includegraphics[width=8.6cm]{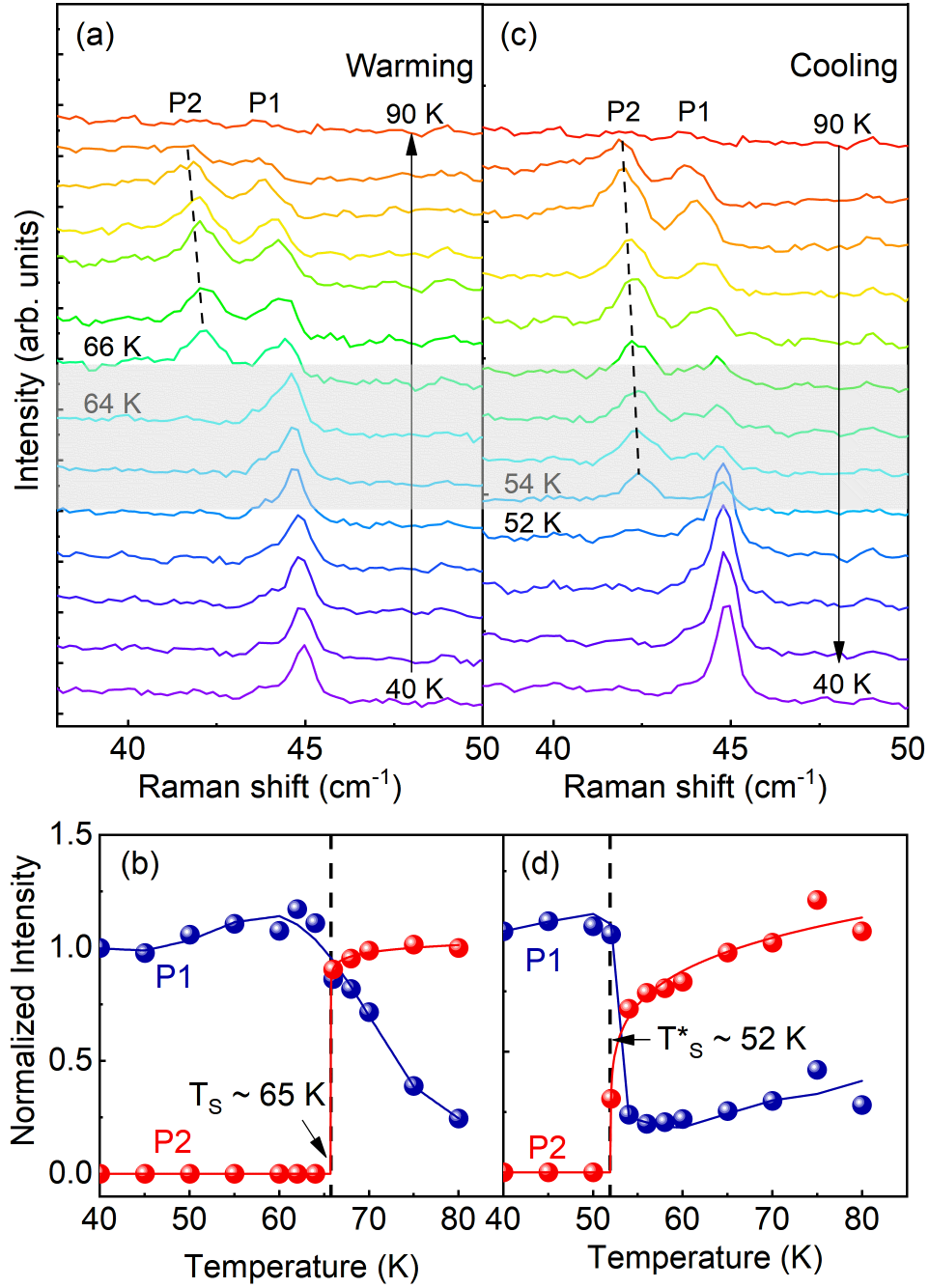}
 \caption{Thermal hysteresis behavior of Cs-related phonon modes. (a) Raman spectra of the Cs-related modes with increasing temperatures. A new mode P2 emerges at around 42 cm$^{-1}$. The normalized integrated intensities of P1 and P2 modes are summarized in panel (b).  (c-d) Same as that in (a) and (b) but with decreasing temperatures, suggesting a structural phase transition at a relatively lower temperature $\sim$52 K.}
\end{figure}

Below, we show that three distinct domains that are oriented at approximately 120$^{\circ}$ to each other can clearly be seen (Fig.~2c-2f). According to equation (1), $I_{A_{g}}$ has a maximum or a minimum when the polarization of incident light is along the principal axis (a/b-axis). For instance, as shown in Fig.~2c, the in-plane principal axis can be determined to be along the approximately 0$^{\circ}$ and 90$^{\circ}$. By performing polarization angle dependent measurements in different regions of one single crystal (the details are given in the Supplemental Material), three distinct domains can be clearly seen and the in-plane principal axis between regions are rotated by approximately 120$^{\circ}$ with respect to each other (Fig.~2c-2e). This characteristic angle indicates that the three domains are formed due to a specific reason but not randomly.

To understand the breaking of $C_6$ rotational symmetry and the special angle of 120$^{\circ}$ between domains, we demonstrate that the CDW phase is a 2c staggered order phase with neighboring layers having a relative $\pi$ phase shift (Fig.~2g-2i). Let us suppose that the Layer1 of the three domains has an ISD pattern with the V-hexagon located at the center of the unit cell. Considering a $\pi$ phase shift translation between adjacent layers, Layer2 has six different ways to deform, namely, a $\pi$ phase shift translation of the V-hexagon along 0$^{\circ}$, 60$^{\circ}$, 120$^{\circ}$, 180$^{\circ}$, 240$^{\circ}$, and 300$^{\circ}$. Of these six directions, the shift along 0$^{\circ}$ and 180$^{\circ}$, 60$^{\circ}$ and 240$^{\circ}$, 120$^{\circ}$ and 300$^{\circ}$ gives the same domain, resulting in three distinct domains with their principal axis rotated by 120$^{\circ}$ with respect to each other. This picture can naturally explain both the breaking of $C_6$ rotational symmetry and the formation of these three distinct domains.

The above results suggest that the CDW phase is a 2c staggered order phase at low temperatures. Now the question is how the stacking configuration evolves with temperatures. Given that the Cs-triangular layer lies between the $V_3Sb_5$ layers, the vibration of Cs ions should be sensitive to the changing of stacking configurations. Thus, to explore the temperature evolution of the stacking configurations, we turn to focus on the temperature dependence of the Cs-related modes.

Previous Raman experiments~\cite{Wu2022} have identified the P1 phonon mode located at $\sim$45 cm$^{-1}$ as a Cs-related mode and we display the temperature-dependent spectra of this mode in Fig.~3a and 3b. Upon warming, the intensity of the Cs-related mode P1 remain nearly constant and then drops abruptly above 65 K. Accompanying the suppression of the P1 mode, a new mode P2 appears at around 42 cm$^{-1}$. The temperature dependence of the P1 and P2 mode's intensities are summarized in Fig.~3b, from which a transitionlike behavior is observed with a critical temperature T$_{s} \sim$ 65 K. The temperature-dependent evolutions of the intensity of P1 and P2 modes remind us that there exists a spectral transformation from P1 to P2 modes. Considering also the similar energy of these two modes, we can identify that the P2 mode, same as P1 mode, is also a Cs-related mode.

\begin{figure}[t]
\centering
\includegraphics[width=8.6cm]{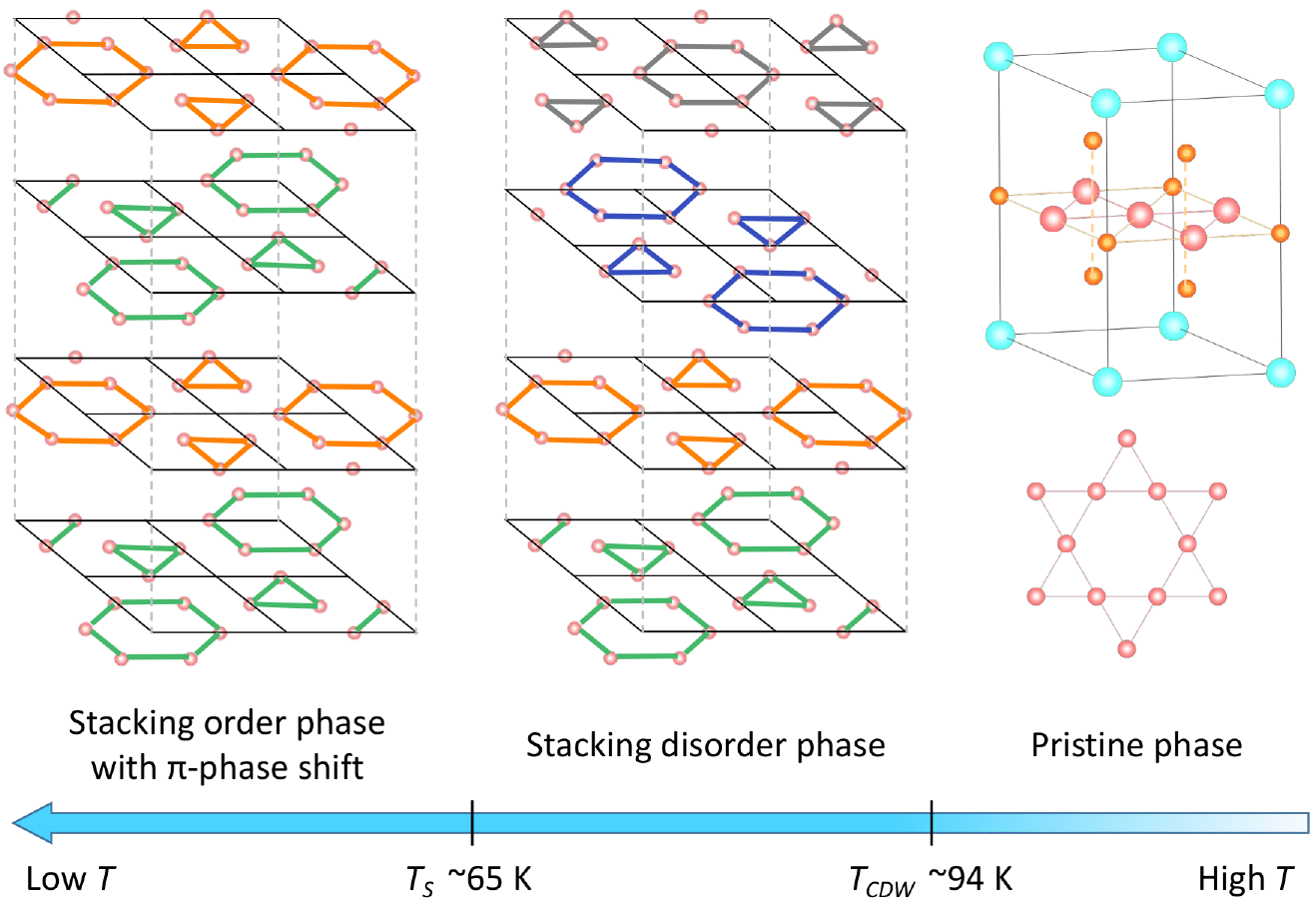}
 \caption{Schematic diagram of phase transitions in CsV$_{3}$Sb$_{5}$.}
\end{figure}

The emergence of a new Cs-related mode indicates a structural phase transition at approximately 65 K, which can be attributed to either an in-plane or an out-of-plane modulations. Having in mind that neither the Cs-triangular layers nor the $V_3Sb_5$ layers exhibit further in-plane deformation in this temperature range~\cite{Nie2022,Stahl2022, Xiao2023} and that the $\pi$ phase shift along different directions (Fig.~2) has the same energy, we propose that the transition at $\sim$65 K is a structural phase transition along the c axis, namely, a stacking order-disorder phase transition from a $\pi$ phase stacking ordered state to a stacking disorder state, as shown in Fig.~4.

To verify the proposal, one can examine the thermal hysteresis behavior of this transition since such an order-disorder phase transition should depend on the history of the stacking configurations. In Fig.~3c, we present the Raman spectra of the P1 and P2 modes with decreasing temperatures. The temperature evolution of these two modes' intensities (Fig.~3d) also exhibits a transitionlike behavior but at a relatively lower temperature T$^*_{s} \sim$ 52 K, suggesting a thermal hysteresis behavior. The thermal hysteresis behavior clearly confirms that the observed structural phase transition at $\sim$65 K is of first-order type.

The first-order structural transition observed here should be related to the anomalies reported by other measurements~\cite{Ratcliff2021,Xiang2021,Chen2021,Stahl2022,Chen2022}. However, it's strange why such a first-order transition has not been reported by previous specific heat measurements~\cite{Ortiz2020} where a broad bump but not a $\lambda$ shape can be observed. In fact, the bump behavior is very similar to that observed in the Ammonium perchlorate NH$_4$ClO$_4$ where a rotational order-disorder transition occurs for the NH$_4^+$ ions~\cite{Brown1989}. The observations suggest that the transition at T$_{s}$ is not a conventional order to order phase transition but an order to disorder transition in which the total release of entropy takes place over a large temperature range due to the existence of short range orders above T$_{s}$. The order-disorder transition proposed here is also supported by the unusual increase of the phonon linewidth~\cite{SuppleM}, the coexistence of the 2c- and 4c-CDW phase below 94 K~\cite{Xiao2023} and the IXS experiments~\cite{Subires2023}. All of these allow us to conclude that the transition at T$_{s}$ is an order-disorder one.

In summary, through systematic polarization- and temperature-dependent Raman measurements, we reveal that the stacking degree of freedom plays a crucial role in determining the properties of CsV$_{3}$Sb$_{5}$. The splitting of the E$_{2g}$ mode and the anisotropic phonon behavior undoubtedly comfirm the breaking of the $C_6$ rotational symmetry. We further observe three distinct domains with their principle axis rotated by $\sim$120$^{\circ}$ to each other. The observations enable us to determine that the CDW phase is a 2c staggered order phase with neighboring layers having a relative $\pi$ phase shift. We further discover a first-order structural phase transition at around 65 K, which can be attributed to a stacking order-disorder phase transition evidenced by the thermal hysteresis behavior of a Cs-related phonon mode. The above results highlight the importance of the stacking degree of freedom in CsV$_{3}$Sb$_{5}$, which add strong constraints on the interpretation of other macroscopic measurements since different stacking configurations or domains may coexist in a single crystal. Moreover, the stacking fault found here offers structural insights to comprehend the pressure/doping induced double superconducting domes and charge instability.

\begin{acknowledgments}
This work was supported by the National Science Foundation of China (No. 12104491, No. U1932215 and No. 12274186), the National Key Research and Development Program of China (No. 2022YFA1402704), the Strategic Priority Research Program of the Chinese Academy of Sciences (No. XDB33010100), and the Synergetic Extreme Condition User Facility (SECUF).
\end{acknowledgments}

\end{document}